\documentclass[acmconf,screen]{acmart}
\usepackage{makecell}

\AtBeginDocument{%
  \providecommand\BibTeX{{%
    \normalfont B\kern-0.5em{\scshape i\kern-0.25em b}\kern-0.8em\TeX}}}

\usepackage{amsmath}
\usepackage{algorithm2e}

\newtheorem{conjecture}{Conjecture}
\newtheorem{setting}{Setting}

\newcounter{fig}

\begin{document}

\title{Optimal Multi-Dimensional Auctions: Conjectures and Simulations}


\author{Alexey Kushnir}
\email{akushnir@andrew.cmu.edu}
\affiliation{%
	\institution{Tepper School of Business, Carnegie Mellon University}
	\streetaddress{4765 Forbes Ave}
	\city{Pittsburgh}
	\state{Pennsylvania}
	\country{USA}
	\postcode{15213}
}
\author{James Michelson}
\email{jamesmic@andrew.cmu.edu}
\affiliation{%
  \institution{Department of Philosophy, Carnegie Mellon University}
  \streetaddress{5000 Forbes Ave}
  \city{Pittsburgh}
  \state{Pennsylvania}
  \country{USA}
  \postcode{15213}
}


\renewcommand{\shortauthors}{Kushnir and Michelson}

\begin{abstract}
  We explore the properties of optimal multi-dimensional auctions in a model where a single object of multiple qualities is sold to several buyers. Using simulations, we test some hypotheses conjectured by Belloni et al. \cite{belloni2010multidimensional} and Kushnir and Shourideh \cite{KuSh2022}. As part of this work, we provide the first open-source library for multi-dimensional auction simulations written in Python.
\end{abstract}




\maketitle

\section{Introduction}
\label{sec:intro}

The mechanism design literature has a good understanding of revenue-maximizing auctions in one-dimensional settings. In his seminal paper on optimal auctions, Myerson \cite{myerson1981optimal} analyzed the problem of selling a single object to multiple-buyers whose values are one-dimensional and independently distributed. He showed that the object should be optimally allocated to a buyer with the highest ``virtual value" whenever it is non-negative. The virtual value is a function of buyer's value that depends only on the buyer's type distribution. It also determines buyer's reserve price and, hence, the set of buyer's types that do not receive the object, i.e., the exclusion region. 

The analysis of the revenue-maximizing multi-dimensional auctions has proved to be much more difficult. Most papers in the literature focus on the problem of a monopolist offering several objects to buyers when the cost function is separable across buyers.\footnote{One of few papers that studies optimal mechanisms for selling multiple objects to multiple buyers is by Armstrong \cite{armstrong2000optimal}. He showed that in a simple setting with a binary distribution of types the analysis proves to be very difficult. The optimal auction takes one of two formats: either objects are sold independently, or a degree of bundling is introduced in the sense that the probability a bidder wins one object is increasing in her value for the other.} In this case, buyers can be dealt one-by-one and the problem effectively becomes how to sell several objects to a single buyer. Armstrong \cite{armstrong1996multiproduct} showed that in almost any optimal auction, it is optimal not to sell with positive probability. This property is considered to be general in the literature. Rochet and Chone \cite{rochet1998ironing} and Manelli and Vincent \cite{manelli2007multidimensional} show that it is applicable to a wide range of multi-dimensional settings (see also \cite{daskalakis2015multi}). These papers also argue that optimal selling mechanism are quite complex, often requiring randomly assigning objects to buyers.

The pioneering work Belloni et al. \cite{belloni2010multidimensional} challenged the above papers suggesting that the optimal multi-dimensional mechanisms with  \textit{multiple buyers} might have quite different properties. They consider a model where a single object of multiple quality grades is sold to multiple buyers who have multi-dimensional types. Using simulations, they show that some optimal mechanisms sell with probability one. Also, they show that the class of ``exclusive buyer mechanisms'' performs quite well relative to the numerical optimal mechanisms in their simulations. These mechanism are simple in the sense that they do not not involve randomization and can be implemented via a modified second-price auction (in dominant strategies). In this paper, we revisit Belloni et al. \cite{belloni2010multidimensional}'s results to verify several broad messages from their work. 
\begin{enumerate}
	\item [1.] \textit{The optimal auction revenue can be well approximated by the optimal exclusive buyer mechanism.}
	\item [2.] \textit{In some multi-buyer settings the measure of buyers' types that receive no object (the exclusion region) in the optimal auction has measure zero in contrast to the corresponding single-buyer setting.}
\end{enumerate}
We show via simulations that there are environments where both statements do not hold. These results lead us to propose a new conjecture that was first suggested by Kushnir and Shourideh \cite{KuSh2022} in a different setting where there are two objects need to be allocated to two buyers (as in \cite{armstrong1996multiproduct}). 
\begin{enumerate}
	\item [3.] \textit{The exclusion region of the optimal allocation is independent of the number of buyers.}
\end{enumerate}
Overall, our simulations allow us to explore interesting properties of multi-dimensional optimal auctions in the case when an auctioneer wants to sell one object of different quality grades to multiple buyers. To accomplish this we develop an open-source software library written entirely in Python which enables researchers to establish new general properties of optimal multi-dimensional mechanisms.

The paper proceeds as follows. Section \ref{sec:model} introduces the model. We outline how we approach the simulation results in Section \ref{sec:simulations}. In Section \ref{sec:conj}, we formally present our main conjectures. These conjectures are tested using several settings in Section \ref{sec:results}. Section \ref{sec:conclusion} concludes.

\section{Model}
\label{sec:model}

We consider a setting similar to the one in Belloni \cite{belloni2010multidimensional}. There is one object that needs to be allocated to $N$ buyers. The object can be tailored to buyers' needs with $j=1,...,J$ possible quality grades. With slight abuse of notation, we denote $N=\{1,...,N\}$ and $J=\{1,...,J\}$.

Each buyer $i\in N$ has a multi-dimensional type $v^i=(v^i_1,...,v^i_J)$ that determines the buyer's willingness to pay for the various quality grades of the object. We assume that buyer's types are distributed according to $F$ with support $V\equiv\Pi_{j\in J}[\underline{v}_j,\overline{v}_j]$. The valuations are independently distributed across buyers, but the valuations for different quality grades could be potentially correlated for each buyer. Also, $v^{-i}$ and $F^{-i}(v^{-i})$ refer the vector of buyers' valuations except buyer $i$ and the distribution of buyers' valuations except buyer $i$. We also assume that buyer's utility is linear. If we denote the probability that buyer $i$ receives the object of grade $j$ as $q^i_j$ and needs to pay $m_i$, buyer $i$'s utility equals
$$
u^i=\sum^J_{j=1}v_j^iq_j^i-m^i.
$$
The costs of producing the object of quality $j$ is constant and denoted $c_j$. When marginal costs are constant one could consider an equivalent model with zero marginal costs by replacing $v_j-c_j$ with a shifted distribution of $v_j$. We keep Belloni et al. \cite{belloni2010multidimensional}'s notation with non-zero marginal costs for easy comparison.

Using the direct revelation principle, we restrict our attention to direct mechanisms with allocation $q=(q^1,...,q^N)$ and payment $m=(m^1,...,m^N)$, where for each $i\in N$
\begin{align*}
&q^i:V^N\rightarrow \Delta(J),\hspace{0.5cm} m^i:V^N\rightarrow \mathbb{R}.
\end{align*}
We also denote interim probability that buyer $i$ is awarded the object of quality $j$ as $Q^i_j(v^i)=\int_{V^{N-1}}q_j^i(v^i,v^{-i})dF^{-i}(v^{-i})$ and interim payments as $M^i(v^i)=\int_{V^{N-1}}m^i(v^i,v^{-i})dF^{-i}(v^{-i}).$ Overall, buyer $i$'s utility from reporting type $\widehat{v}^i$ when her true type is $v^i$ equals
$$
U^i(\widehat{v}^i,v^i)=\sum^J_{j=1}v_j^iQ_j^i(\widehat{v}^i)-M^i(\widehat{v}^i).
$$
A mechanism $(q,m)$ is \textit{incentive compatible} if truthtelling is a Bayes-Nash equilibrium, i.e., 
\begin{equation}
U^i(v^i,v^i)\geq U^i(\widehat{v}^i,v^i) \text{\hspace{0.5cm} for all  } \widehat{v},v\in V \text{ and } i\in N.\tag{IC}\label{IC}
\end{equation}
Also, a mechanism $(q,m)$ is \textit{individual rational} if 
\begin{equation}
	U^i(v^i,v^i)\geq 0 \text{\hspace{0.5cm} for all } v\in V\text{ and }i\in N.\tag{IR}\label{IR}
\end{equation}

The seller's problem can be formulated as follows

\vspace{1mm}
\begin{align*}
&\max_{q,m} \int_{V^N}\sum_{i\in N}\left[m^i(v)-\sum_{j=1}^J c_jq_j^i(v)\right] dF(v)\\
&\text{s.t.}\left\{\begin{array}{l}
	(\ref{IC}),(\ref{IR}), \text{ and }\\
	\sum_{i\in N}\sum_{j\in J} q_j^i(v)\leq 1, q_j^i(v)\geq 0 \text{\hspace{0.5cm} for all } v\in V,i\in N,j\in J.
\end{array}\right.
\end{align*}

\vspace{3mm}
\noindent This problem can be rewritten in a more convenient way. First, we restrict ourselves to symmetric allocations\footnote{This is only for the simplicity of exposition. We do consider asymetric settings in our simulations.}. Hence, we drop index $i$ from $Q^i_j$, $M^i$, $U^i$, $v^i$, $q^i_j$, and $m^i$. With some abuse of notation, we also denote buyer's utility from truthtelling as $U(v)\equiv U(v,v)$. Then, the incentive compatibility constraints can be rewritten as
\begin{equation}
	U(v)-U(\widehat{v})\geq \sum_j Q_j(\widehat{v})(v_j-\widehat{v}_j) \text{\hspace{0.5cm} for all  } \widehat{v},v\in V,
	\tag{ICC}\label{ICC}
\end{equation}
and the individual rationality constraints can be rewritten as
\begin{equation}
	U(v)\geq 0 \text{\hspace{0.5cm} for all  } v\in V.
	\tag{IRR}\label{IRR}
\end{equation}
There is no reason to leave any rents to the buyer with the lowest type. Hence, $U(\underline{v})=0$. At the same time, the incentive compatibility constraints imply that $U$ is convex and non-decreasing and, hence, $U(\underline{v})=0$ guarantees that all individual rationality constraints are satisfied.

Finally, using the seminal result due to Border \cite{border1991implementation}, the feasibility constraints can be rewritten in terms of interim probabilities of buyers getting the object as
\begin{equation}
\left\{\begin{array}{l}
	Q_j(v)\geq 0 \text{\hspace{0.5cm} for all } v\in V,j\in J\\
	N \int_A\sum_{j\in J}Q_j(v)dF(v)\leq 1-\left(\int_{V\backslash A}dF(v)\right)^N \text{ for all } A\subset V
\end{array}\right..
\tag{B}\label{B}
\end{equation} 
Finally, the optimization problem can be rewritten in terms of optimization over $Q$ and $U$.
The seller's problem can be formulated as follows
\begin{equation}
	\left\{\begin{array}{l}
		OPT_*=\max_{Q,U} \int_{V}\sum_{j\in J}(v_j-c_j)Q_j(v)-U(v)dF(v)\\[2mm]
		\text{subject to  }(\ref{B}),\,(\ref{ICC}),\,U(\underline{v})=0.
	\end{array}\right..
\tag{$P^*$}\label{P*}
\end{equation}

\section{Simulations}
\label{sec:simulations}

We adopt and improve the original finite-dimensional approximation algorithm of Belloni et al. \cite{belloni2010multidimensional} by focusing on local and downward-sloping incentive-compatibility constraint (ICC) violations. Although these local constraints are often violated in this approximate setting, we drastically reduce the number of times \textit{all} incentive-compatible constraints need to be checked.

In order to approximate an optimal solution to \ref{P*}, we discretize the type space $V$\footnote{This presentation mirrors Belloni et al. \cite[\S 3]{belloni2010multidimensional}.}. Let $T$ denote a positive integer that controls the granularity of the discretization. For each $j \in J$, let $V_T(j)$ denote the discretization of the interval $[\underline{v}_j,\overline{v}_j]$ given by $V_T(j) = \{\underline{v}_j, \underline{v}_j + \epsilon, \underline{v}_j + 2\epsilon, \dots, \overline{v}_j\}$ where $\epsilon = \min_{j \in J} \{(\overline{v}_j - \underline{v}_j) / T\}$. Our discretized version of the type space $V$ is given by $V_T := \prod_{j \in J} V_T(j)$. Furthermore, we define a probability density function on $V_T$ by setting $\hat{f}(v) = f(v) / (\sum_{t \in V_T} f(t))$. We thus obtain a linear program which is a finite-dimensional approximation of \ref{P*} for each $T > 0$ by replacing $V$ with $V_T$.

\begin{figure}[t]
    \begin{center}
    \includegraphics[width=0.3\textwidth]{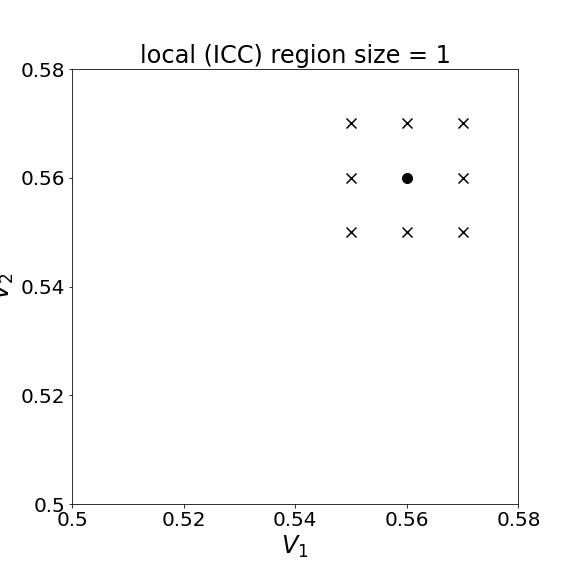}
    \includegraphics[width=0.3\textwidth]{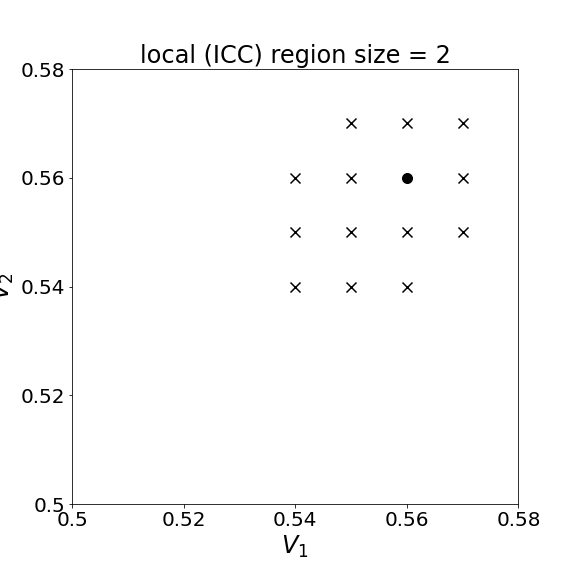}
    \includegraphics[width=0.3\textwidth]{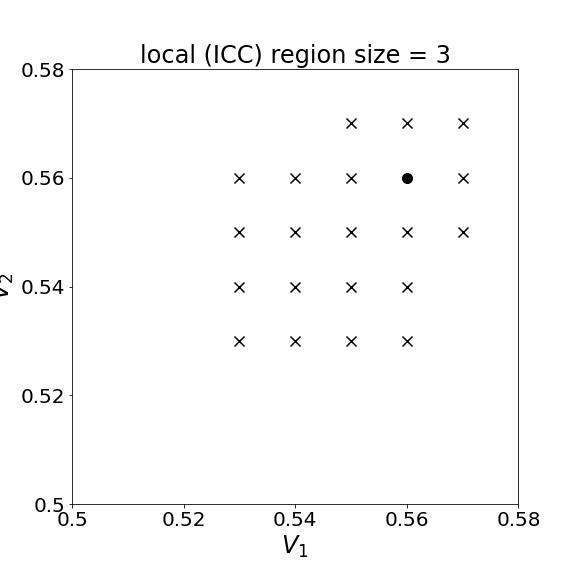}
    \end{center}
    
    \vspace{1mm}
    \raggedright{\small {\sc Figure \refstepcounter{fig}\thefig\label{fig:sim1}:} We iteratively grow the local region of the discretized type space checked for downwards-sloping constraint violations. Notice that immediately adjacent (ICC) constraints are always checked ($\times$) when the local region increases in size around a point ($\bullet$).}
\end{figure}

Belloni et al. \cite[Algorithm 11]{belloni2010multidimensional} use a plane-putting algorithm which works with a randomly chosen subset of incentive-compatibility (ICC) and Border (B) constraints at each iteration. Belloni et al. \cite[Lemma 10]{belloni2010multidimensional} provide an efficient reduction in the growth in $T$ of the Border constraints (B) from $O(2^{T^J})$ to $O(T^J \log(T^J))$. We adopt their solution to checking (B) constraints; however, our approach to checking (ICC) constraints involves iteratively growing the `local' region of the type space around each point $v$ in the distcretized set of types $V_T$. We do two things. First, all the immediately adjacent points in the discretized type space are always checked for incentive compatbility. Secondly, downwards-sloping points in the discretized type space are also checked. Furthermore, the downwards-sloping region of the type space grows until all (ICC) constraints are ultimately satisfied. This procedure is illustrated in Figure \ref{fig:sim1}. Thus, for a fixed-size local region aroud each point in the discretized type space, we first satisfy \textit{local} (ICC) and (B) constraints as in the iterative plane-cutting algorithm of Belloni et al. \cite{belloni2010multidimensional}. Then we run the separation oracle with \textit{all} (ICC) and (B) constraints. We then restart the solver with any previously violated constraints, this time increasing the size of the local region around each point in the discretized type space. This procedure iterates until no constraints are violated. This modified version of Belloni et al.'s \cite{belloni2010multidimensional} algorithm is described in Algorithm \ref{alg:1}.

\begin{figure}[ht]
  \centering
  \begin{minipage}{.7\linewidth}
    \begin{algorithm}[H]
    \caption{Iterative plane-cutting algorithm with local and downwards-sloping (ICC) constraints}\label{alg:1}
    \SetAlgoLined
    $L = 1, S = \emptyset, A = \emptyset, \overline{OPT} = \infty$\;
    \texttt{violated\_any\_icc} $\gets $ TRUE\; 
    \While{\texttt{violated\_any\_icc}}{
        \texttt{violated\_local\_icc} $\gets$ TRUE\; 
        \While{\texttt{violated\_local\_icc}}{
          $k = 1, A^k = A, S^k = S$\;
          Solve the linear program associated with $S^k$. Let $OPT^k$ denote the optimal value.\;
          Solve the separation oracle using only local (ICC) constraints in region $L$. Let $A^k$ donate all violated local (ICC) and (B) constraints.\;
          \If{$A^k = \emptyset$}{
            \texttt{violated\_local\_icc} $\gets$ FALSE\;
            Break\;
          }
          Select a subset $I^k \subset S^k$ of inactive (ICC) and (B) constraints\;
          \eIf{$OPT^k < \overline{OPT}$}{
            $S^{k+1} \gets (S^k \setminus I^k) \cup A^k \cup A$\;
            $\overline{OPT} \gets OPT^k$\;
          }{
            $S^{k+1} \gets S^k \cup A^k \cup A$\;
          }
          $k \gets k+1$\;
        }
        Solve the separation oracle using all (ICC) constraints. Let $A^*$ donate all violated (ICC) constraints.\;
        \If{$A^* = \emptyset$}{
            \texttt{violated\_any\_icc} $\gets$ FALSE\;
            Break\;
        }
        $A \gets A \cup A^*$\;
        $S \gets S^k$\;
        $L \gets L + 1$\;
    }
    \end{algorithm}
  \end{minipage}
\end{figure}

Our algorithm is written in Python 3.10 and uses Google's open source linear programming solver `GLOP' available in their \texttt{or-tools} package \cite{ortools}. Algorithm \ref{alg:1} recovers similar optimal values to that of Belloni et al. \cite{belloni2010multidimensional}.

\section{Conjectures}
\label{sec:conj}

The principal message of Belloni et al. \cite{belloni2010multidimensional} is that the revenue maximizing mechanisms in multiple-buyer settings could have quite different qualitative features compared to the revenue maximizing mechanisms in the single-buyer case. We investigate this message through several conjectures.

The mechanism design literature analyzing single-buyer settings (see Armstrong \cite{armstrong1996multiproduct}, Rochet and Chone \cite{rochet1998ironing}, Manelli and Vincent \cite{manelli2007multidimensional}) argues that optimal revenue-maximizing allocation mechanisms could be quite complicated. The optimal allocation mechanism is characterized by bunching regions, i.e., different buyers' types receiving the same allocation of objects. In some cases, the optimal allocation mechanism has to be random, i.e., buyers receive objects only with some probability (see also Daskalakis et al. \cite{daskalakis2017strong}). In contrast, Belloni et al. \cite{belloni2010multidimensional} propose the class of ``exclusive buyer mechanisms'' that performs quite well relative to the numerical optimal mechanisms in their simulations. In an exclusive buyer mechanism, buyers compete in the second-price auction for the right to be the only buyer to get the object. The winner then chooses quality $q_j$ of the object to acquire at price $p_j$. In other words, denote buyer $i$'s function of its value 
$$
\beta^i=\max(v_1^i-p_1,...,v_J^i-p_J).
$$ 
In the exclusive buyer mechanism, buyer $i$ with the highest value $\beta^i$ wins the object and chooses the object quality grade $j$ for price $\max_{l\neq i}(v^l_j,p_j)$.\footnote{Note that Belloni et al. \cite{belloni2010multidimensional} used a slightly different language to define an ``exclusive buyer mechanism.'' They consider only the case with two objects and assumed that the second-price auction could be augmented with a reserve price. At the same time, one of the object qualities is allocated at no additional costs in their original formulation. Both formulations are equivalent.} 
\begin{conjecture}
	The optimal auction revenue can be well approximated by the optimal exclusive buyer mechanism.
	\label{conj:1}
\end{conjecture}
A further important insight by Belloni et al. \cite{belloni2010multidimensional} is that for some multi-dimensional settings the optimal selling mechanism with multiple buyers has no exclusion region. This is in sharp contrast to Armstrong \cite{armstrong1996multiproduct} and Rochet and Chone \cite{rochet1998ironing} who show that it is optimal not to sell with positive probability in multi-dimensional settings with a single-buyer. 
\begin{conjecture}
	In some multi-buyer settings the measure of buyers' types that receive no object (the exclusion region) in the optimal auction has measure zero in contrast to the corresponding single-buyer setting.
	\label{conj:2}
\end{conjecture}
We will see below (see Section \ref{sec:results}) that we are able to reject both conjectures. These conclusions, together with other simulations that we performed, bring us to investigate another conjecture that connects the exclusion region in optimal mechanisms for the single-buyer and multiple-buyer settings. This conjecture was first proposed by Kushnir and Shourideh \cite{KuSh2022} in a different setting where there are two objects need to be allocated to two buyers (as in \cite{armstrong1996multiproduct}). 
\begin{conjecture}
	The exclusion region of the optimal allocation is independent of the number of buyers $N=1,2,3,...$
	\label{conj:3}
\end{conjecture}
In the next section, we provide simulation results addressing these conjectures. 

\section{Results}
\label{sec:results}
To study the above conjectures, we consider several settings. All settings concern either single or multiple buyers ($N=1,2,3$) who want to buy a single-good of two possible quality grades ($J=2$). In all simulations that follow each dimension of the type space $V$ is discretized into $T=20$ intervals.

\begin{figure}[t]
	\begin{center}
	\includegraphics[width=0.3\textwidth]{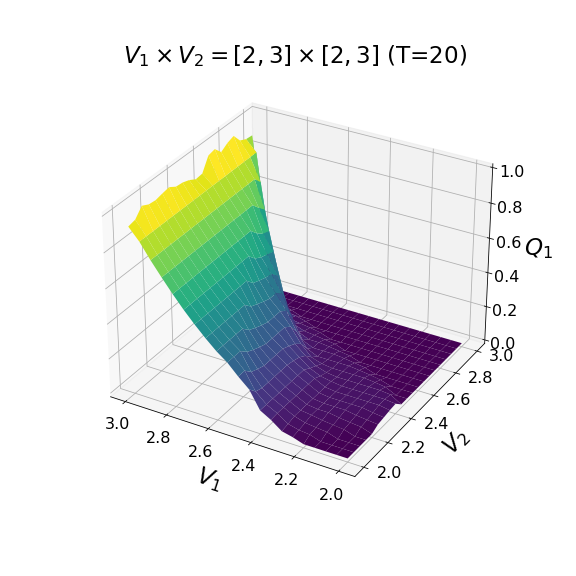}
	\includegraphics[width=0.3\textwidth]{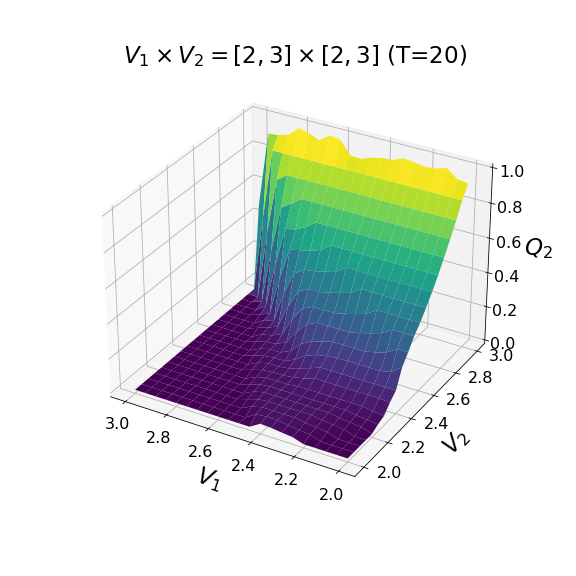}
	\includegraphics[width=0.3\textwidth]{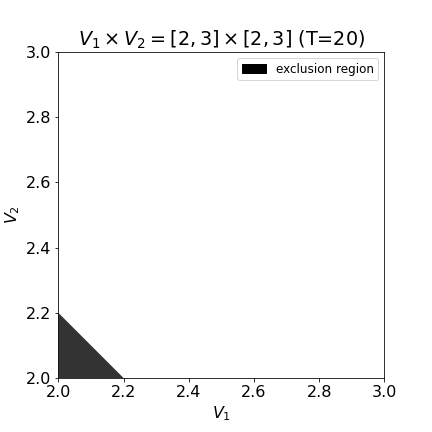}
	\end{center}
	
	\vspace{1mm}
	\raggedright{\small {\sc Figure \refstepcounter{fig}\thefig\label{fig:1}:} The optimal auction in Setting \ref{set:1}. The first figure and the second figure show the interim allocation probabilities for quality 1 ($Q_1$) and quality 2 ($Q_2$). The third figure depicts the exclusion region -- the set of buyer values that receive object with zero probability.}
\end{figure}

Let us consider first a setting where buyer types are uniformly distributed on $[2,3]\times[2,3]$. This is an extension of the environment that was analyzed by Pavlov \cite{pavlov2011optimal} for the optimal mechanism in the single-buyer case.
\begin{setting}
	Uniform distribution on $[2,3]\times[2,3]$, costs $c_1=0$, $c_2=0$, $N=2$ buyers, and $J=2$ qualities.
	\label{set:1}
\end{setting}
Figure \ref{fig:1}  presents the results of our simulations. The first graph shows the probability of obtaining object of quality 1, the second graph shows the probability of obtaining object of quality 2, and the last graphs depicts the exclusion region, i.e. the set of buyer's types that does not receive any object. 

We can observe that the exclusion region in the optimal auction is not rectangular. Hence, the allocation rule cannot be replicated with an exclusive buyer mechanism introduced by Belloni et al. \cite{belloni2010multidimensional}. The optimal auction has different qualitative properties with respect to the exclusion region. Belloni et al. report the optimal auction revenue and the revenue in the optimal exclusive buyer mechanism coincide in their simulated setting (see Tables 2 and 3 in \cite{belloni2010multidimensional}) subject to computational error (i.e., $<10^{-6}$). We also find that, despite qualitative difference, the difference in the revenue of optimal auction and the revenue of the optimal exclusive buyer mechanism is around one percent. 

\begin{figure}[t]
	\begin{center}
	\includegraphics[width=0.3\textwidth]{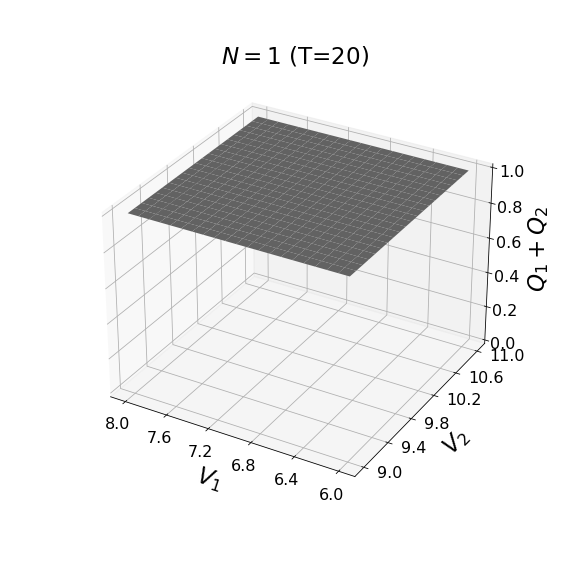}
	\includegraphics[width=0.3\textwidth]{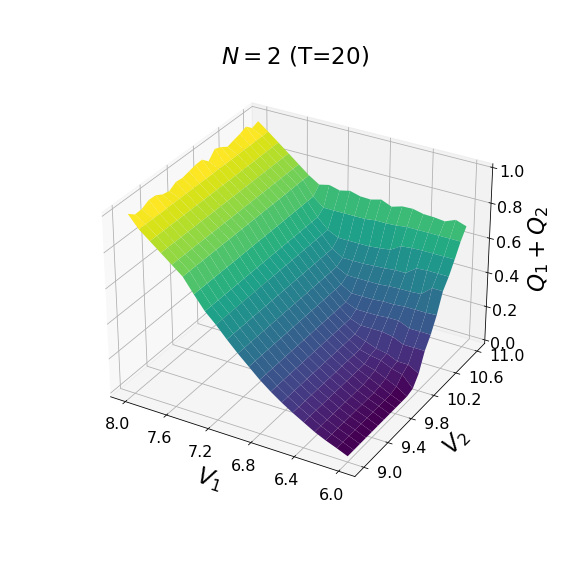}
	\end{center}

	\vspace{1mm}
	\raggedright{\small {\sc Figure \refstepcounter{fig}\thefig\label{fig:2}:} The optimal mechansims for Setting \ref{set:2}. The first graph shows the allocation for quality 1 ($Q_1$) and quality 2 ($Q_2$) for the single-buyer case. The second figure shows the allocation for quality 1 ($Q_1$) and quality 2 ($Q_2$) for the two-buyer case.}
\end{figure}

To analyze Conjecture \ref{conj:2} we consider the following setting.
\begin{setting}
	Uniform distribution on $[6,8]\times[9,11]$, costs $c_1=0.9, c_2=5$, $N=1,2$ buyers, $J=2$ qualities
	\label{set:2}	
\end{setting}
The above setting for $N=2$ is the main simulation setting of Belloni et al. \cite[Figure 1]{belloni2010multidimensional}. Using simulations, they discovered that there is no exclusion region in the case of multiple buyers. This is in stark contrast with the findings of Armstrong \cite{armstrong1996multiproduct} and Rochet and Chone \cite{rochet1998ironing} in the single-buyer case. To verify Belloni et al. \cite{belloni2010multidimensional} claim that the nature of the optimal selling mechanisms are different for the single-buyer and multiple-buyer cases, we decided to run simulations in the same setting, but only for $N=1$ buyer. The results for both runs are presented in Figure \ref{fig:2}.
The figure illustrates that it is always optimal to sell the object in both $N=1$ and $N=2$. This is different to the seminal result by Armstrong \cite{armstrong1996multiproduct} and Rochet and Chone \cite{rochet1998ironing}! 

Though this might seem to be a contradictory result, it is important to note that Armstrong \cite{armstrong1996multiproduct} and Rochet and Chone \cite{rochet1998ironing} consider a different setting than \cite{belloni2010multidimensional}. The two early papers consider a problem of selling \textit{two objects} to single buyer with each object having one quality grade. However, Belloni et al. \cite{belloni2010multidimensional} consider the setting with only \textit{one object} that might be of several quality grades. The original Armstrong's result requires the set of possible types to be strictly convex. Clearly, the uniform distribution $[6,8]\times[9,11]$ clearly does not satisfy this condition. Rochet and Chone \cite{rochet1998ironing} requires that the cost function of producing the objects to be smooth (see Theorem 1 in \cite{rochet1998ironing}). This assumption is not satisfied in the setting considered by Belloni et al. \cite{belloni2010multidimensional}.

Armstrong \cite{armstrong1996multiproduct} argues that if the set of possible buyer types is strictly convex, then if the measure of exclusion region is zero, there has to be a unique type that receives zero utility in optimal allocation. Then, if we increase the tariff by $\epsilon$ uniformly among all buyer types, there number of types that will be excluded is proportional to $\epsilon^2$. At the same time, the revenue increases proportional to  $\epsilon$. Hence, a uniform increase of the tariff by $\epsilon$ increases revenue. Armstrong's logic does not hold for the example presented above. The set of types is not strictly convex. In addition, one could notice that Figure \ref{fig:2} suggests that the whole interval $6\times [9,p^*]$ for some $p^*\in [9,11]$ receives the same zero utility in equilibrium. The measure of this interval is zero. However, if one increases tariff by $\epsilon$, one would exclude a mass of types proportional to $\epsilon$ not $\epsilon^2$. This breaks down Armstrong's logic. This is a surprising and novel observation to the best of our knowledge.

During the analysis of the previous conjectures, we noticed that the exclusion region in the optimal auction does not depend on the number of buyers (see Conjecture \ref{conj:3}). Hence, we decided to extensively test this conjecture in numerous simulaton setting including symmetric distributions buyers' types, asymetric distributions of buyers' types, and correlated distributions of buyers' types. Some of these settings are presented below.
\begin{setting}
	We consider several settings all with $J=2$ qualities:
	\begin{enumerate}
		\item [a)] 	Symmetric uniform, truncated normal, and beta\footnote{We use the beta distribution given by probability density function $f(x;a,b)=\frac{\Gamma(a+b)}{\Gamma(a)\Gamma(b)}x^{a-1}(1-x)^{b-1}$ where $a=1,b=2$.} distributions on $[0,1]\times[0,1]$, $c_1=0, c_2=0$, $N=1,2,3$ buyers;
		\item [b)] Non-symmetric distributions (uniform $\times$ beta, uniform $\times$ truncated normal, beta $\times$ truncated normal) on $[0,1] \times [0,1],$ $c_1=0, c_2=0$, $N=1,2,3$ buyers; and,
		\item [c)] Mixture of ($\alpha=\frac{1}{3},\frac{1}{2},\frac{2}{3}$) uniform and $(1-\alpha)$ beta distributions on $[0,1] \times [0,1],$ $c_1=0, c_2=0,$ $N=1,2,3$ buyers.
	\end{enumerate}
\label{set:3}
\end{setting}
The results of these simulations are depicted below in Figure \ref{fig:3}. Every graph of the exclusion region is the same for $N=1,2,3$ buyers. The uniformity of these and other our simulation results across a diversity of settings provides strong support for Conjecture \ref{conj:3}. To the best of our knowledge this is an interesting and novel feature of optimal multi-dimensional auctions that was originally proposed in a working paper by Kushnir and Shourideh \cite{KuSh2022} in a different setting where there are two objects need to be allocated to two buyers (as in \cite{armstrong1996multiproduct}).

Overall, we do not consider simulations as definitive evidence for or against the above conjectures. Rather, we consider our simulations as empirical evidence which should guide future theoretical research in multi-dimensional mechansism design.
\newpage

\begin{figure}[t]
	\begin{center}
	\includegraphics[width=0.3\textwidth]{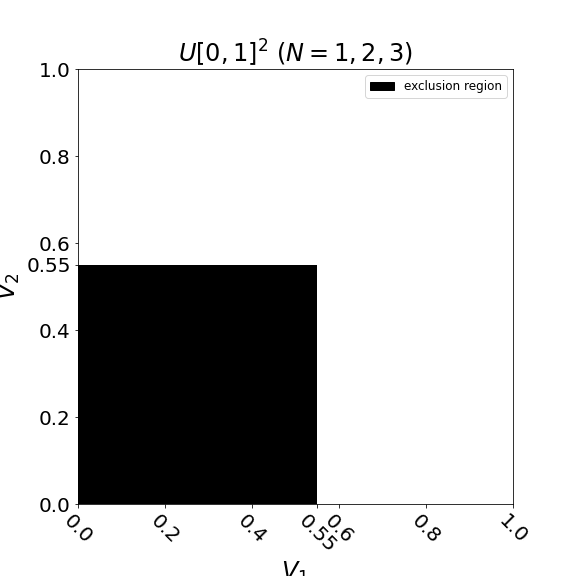}
	\includegraphics[width=0.3\textwidth]{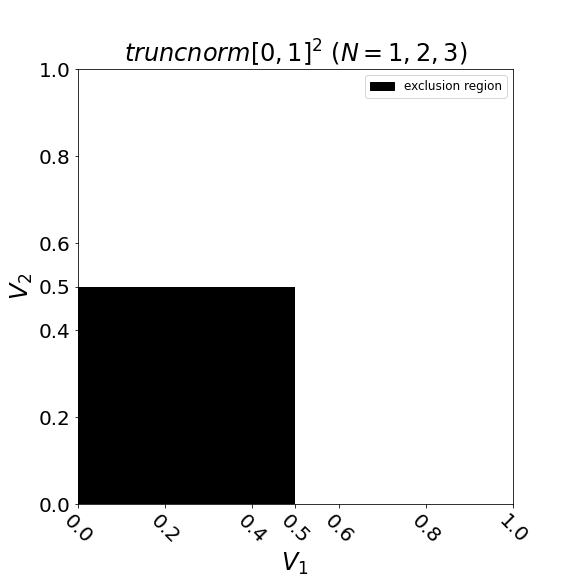}
    \includegraphics[width=0.3\textwidth]{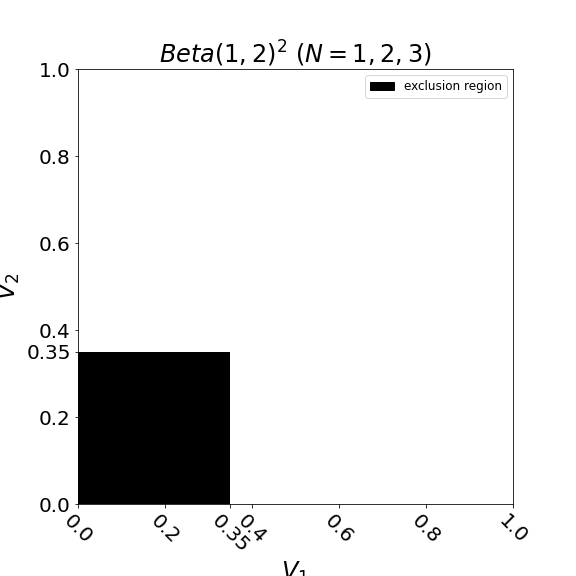}

	\vspace{5mm}
	\includegraphics[width=0.3\textwidth]{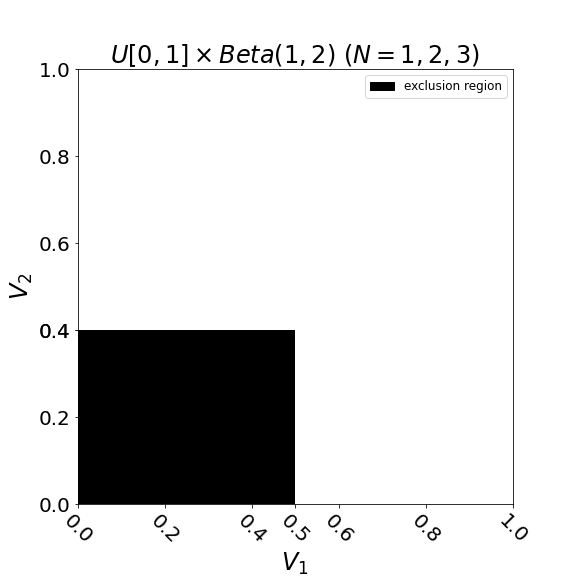}
	\includegraphics[width=0.3\textwidth]{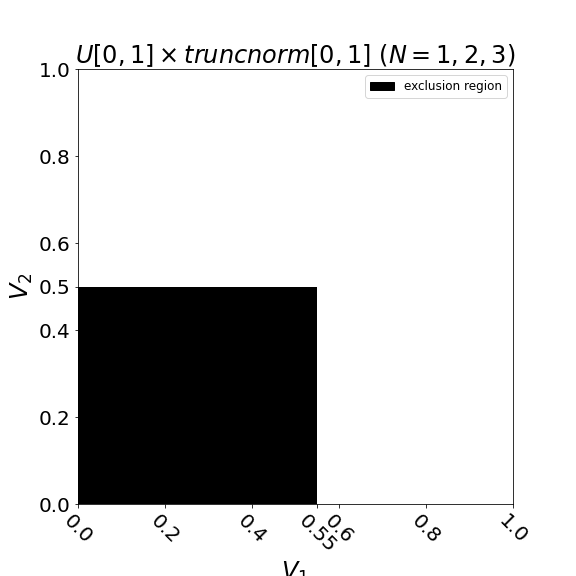}
    \includegraphics[width=0.3\textwidth]{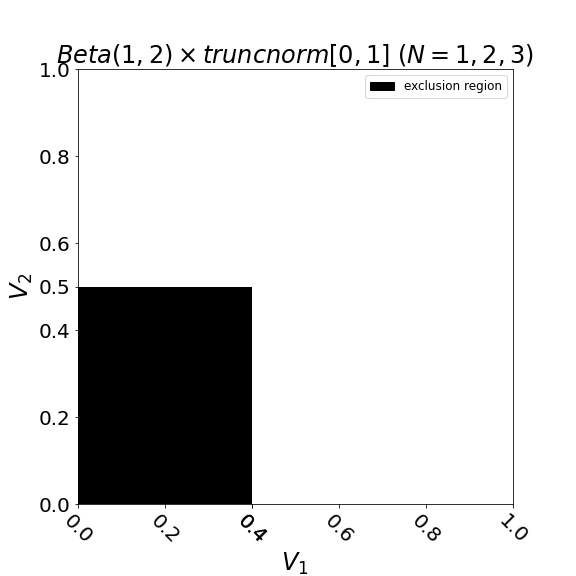}

	\vspace{5mm}
	\includegraphics[width=0.3\textwidth]{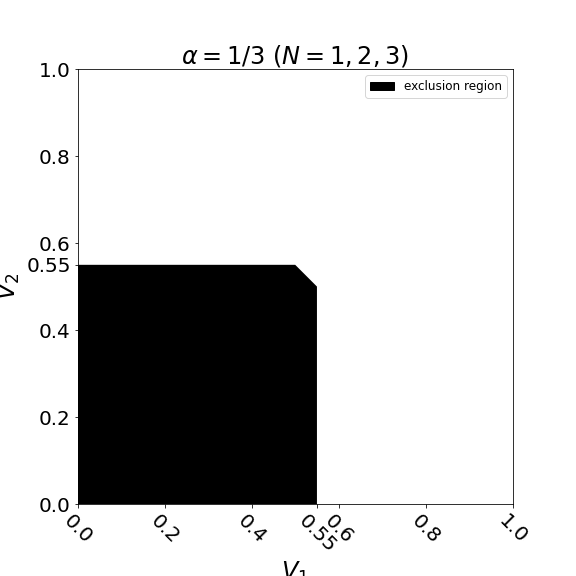}
	\includegraphics[width=0.3\textwidth]{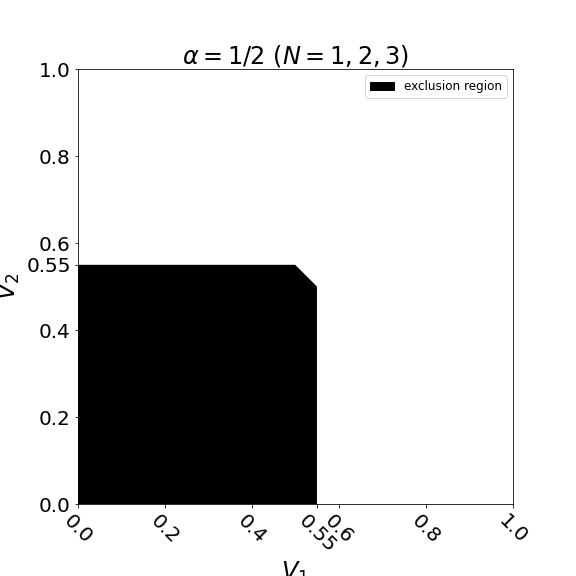}
    \includegraphics[width=0.3\textwidth]{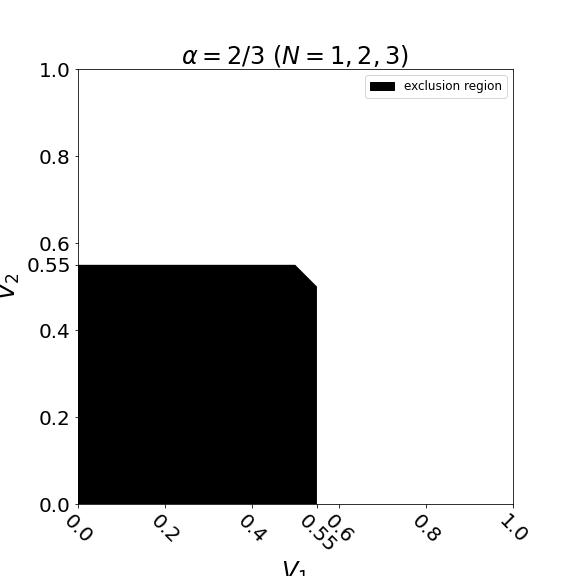}
	\end{center}

	\vspace{5mm}
	\raggedright {\small {\sc Figure \refstepcounter{fig}\thefig\label{fig:3}:} The exclusion regions in optimal auctions corresponding to settings \ref{set:3}. The top figures depict a) the exclusion region for symmetric uniform, truncated normal, and beta distributions with $N=1,2,3$ buyers. The middle figures depict b) the exclusion region for uniform $\times$ beta, uniform $\times$ truncated normal, beta $\times$ truncated normal distributions for $N=1,2,3$ buyers. Lastly, the figures in the bottom row depict c) the exclusion region for the mixture of $\alpha=\frac{1}{3},\frac{1}{2},\frac{2}{3}$ uniform and $(1-\alpha)$ beta distributions for $N=1,2,3$ buyers.}
\end{figure}

\newpage
\pagebreak

\section{Conclusion}
\label{sec:conclusion}
We used simulations to study the properties of revenue maximizing auctions in multi-dimensional settings. We showed that the set of buyer exclusive mechanisms, proposed by Belloni et al. \cite{belloni2010multidimensional}, does not approximate well  the optimal revenue auction in some environments. We also highlight that the absence of an exclusion region in main simulation example in Belloni et al. \cite{belloni2010multidimensional} is not due to the difference in the properties of the optimal auctions in the single-buyer and multiple-buyer case. Instead, Belloni et al. \cite{belloni2010multidimensional}'s setting with only one object of multiple qualities grades is different from classical settings in Armstrong \cite{armstrong1996multiproduct} and Rochet and Chone \cite{rochet1998ironing}, where two objects each having one quality type are allocated. Hence, the classical result about the presence of an exclusion region in single-buyer setting does not apply to Belloni et al. \cite{belloni2010multidimensional} setting. The observation that in multi-dimensional settings the exclusion region in the optimal auctions could be absent is novel in the literature. Our simulations also support the conjecture first proposed by Kushnir and Shourideh \cite{KuSh2022} that the exclusion region does not depend on the number of buyers. 

We treat our simulation results as guidance for future theoretical work. Our next step is to identify conditions under which the exclusion region has measure zero in optimal multi-dimensional auctions in Belloni et al. \cite{belloni2010multidimensional}'s setting. Additionally, we plan to analyze the conditions when the exclusion region does not change with the number of buyers. Furthermore, we have chosen to open-source our numerical library for simulating multi-dimensional auctions to enable other researchers to explore their own conjectures. We hope that other researchers use this library to explore their own hypotheses and conjectures concerning multi-dimensional auctions.

\bibliographystyle{ACM-Reference-Format}
\bibliography{auctions-references}

\end{document}